\documentclass[prl,english,preprintnumbers,amsmath,amssymb,nofootinbib,twocolumn]{revtex4-1}

\pdfoutput=1

\usepackage[latin1]{inputenc}
\usepackage{graphicx}
\usepackage{color}
\usepackage{bbm}
\usepackage{amssymb}
\usepackage{amsmath}

\usepackage{dsfont}

\def\0#1#2{\frac{#1}{#2}}

\def\s0#1#2{\mbox{\small{$ \frac{#1}{#2} $}}}

\newcommand{\Tr}{\mathrm{Tr}}

\newcommand{\E}{\mathrm{e}}
\newcommand{\I}{\mathrm{i}}
\newcommand{\be}{\begin{eqnarray}}
\newcommand{\ee}{\end{eqnarray}}

\newcommand{\nn}{\nonumber }

\newcommand{\zgc}{\mathcal Z}

\usepackage{babel}
\makeatother
\begin{document}

\title{Imaginary polarization as a way to surmount the sign problem\\ in ab
  initio calculations of spin-imbalanced Fermi gases}

\author{Jens Braun$^1$}
\author{Jiunn-Wei Chen$^2$}
\author{Jian Deng$^3$}
\author{Joaqu\'{\i}n E. Drut$^4$}
\author{Bengt Friman$^5$}
\author{Chen-Te Ma$^2$}
\author{Yu-Dai Tsai$^6$}

\affiliation{$^1$Institut f\"ur Kernphysik (Theoriezentrum), Technische Universit\"at Darmstadt,
D-64289 Darmstadt, Germany}
\affiliation{$^2$Department of Physics, National Center for Theoretical Sciences, and Leung Center for Cosmology and Particle Astrophysics, \\
National Taiwan University, Taipei 10617, Taiwan}
\affiliation{$^3$School of Physics, Shandong University, Shandong 250100, People's Republic of China}
\affiliation{$^4$Department of Physics and Astronomy, University of North Carolina, Chapel Hill, NC 27599, USA}
\affiliation{$^5$GSI Helmholtzzentrum f\"ur Schwerionenforschung, D-64291 Darmstadt, Germany}
\affiliation{$^6$Department of Physics, Tsing-Hua University, Hsinchu, Taiwan 300, R.O.C. and \\
Department of Physics, University of California, Berkeley, CA 94720, USA}

\begin{abstract}
From ultracold atoms to quantum chromodynamics, reliable {\it ab initio} studies of strongly interacting fermions
require numerical methods, typically in some form of quantum Monte Carlo. Unfortunately, (non-)relativistic systems
at finite density (spin polarization) generally have a sign problem, such that those {\it ab initio} calculations are impractical.
It is well known, however, that in the relativistic case imaginary chemical potentials solve this problem,
assuming the data can be analytically continued to the real axis. Is this feasible for non-relativistic systems?
Are the interesting features of the phase diagram accessible in this manner? Introducing complex chemical
potentials, for real total particle number and imaginary polarization, the sign problem is avoided in the
non-relativistic case. To give a first answer to the above questions, we perform a mean-field study of the
finite-temperature phase diagram of spin-$1/2$ fermions with imaginary polarization.
\end{abstract}

\maketitle
%
Ultracold Fermi gases provide an accessible and clean environment to study quantum
many-body phenomena~\cite{RevTheory,RevExp}, ranging from Bose-Einstein condensation
(BEC) to Bardeen-Cooper-Schrieffer (BCS) superfluidity. In the dilute limit, where the range
of the interaction is smaller than any other scale, a single parameter $(k^{}_{\rm F} a^{}_{\rm s})^{-1}$,
where $a_{\rm s}$ is the s-wave scattering length and $k^{}_{\rm F}$ the Fermi momentum, describes
the microscopic interactions completely. These are tuned by an external
magnetic field in the presence of a Feshbach resonance.

At large $a^{}_{\rm s}$ (at the crossover between BEC and BCS), these systems display
universal properties. Here, the scale for all physical observables is set solely by $k^{}_{\rm F}$
(or, equivalently, the density), which is the only scale left in the problem. Thus, no obvious small
expansion parameter exists in this limit. This represents a major challenge for theoretical
many-body approaches~\cite{ZwergerBook}. Recently, experiments in this so-called unitary regime have achieved
high precision~\cite{PrecisionExp}, which potentially facilitates benchmarking of
theoretical methods.

In spite of the rapid experimental and theoretical advances, our understanding of
ultracold Fermi gases at unitarity remains incomplete, most notably for the case of spin-imbalanced
systems. For a sufficiently large imbalance, one expects a phase transition
from a BCS-type superfluid to a polarized normal gas.
Such a transition was observed in experiments at MIT and Rice
university~\cite{expasym} and is in accordance with various theoretical studies~\cite{Holland:2001zz,theoasym,revasym}.

Apart from ultracold gases, a better understanding of imbalanced fermion systems is of
great importance also in other research fields. For example, lattice
Monte Carlo (MC) calculations of nuclei~\cite{Epelbaum:2009pd} are confronted with similar problems
in isospin asymmetric nuclei, i.e., nuclei with an unequal number of neutrons and protons.

We shall focus on the $a^{}_{\rm s}\!\to\!\infty$ limit for a spin-imbalanced two-component Fermi
gas at zero and finite temperature. Unlike previous studies~\cite{Holland:2001zz,theoasym,revasym},
however, we consider complex valued chemical potentials of the spin components, $\mu_{\uparrow}$ and $\mu_{\downarrow}$.
In {\it ab initio} MC calculations one can thus avoid
the sign problem, which impedes studies of spin-polarized  systems for real $\mu$.

This approach parallels that of a purely imaginary $\mu$ in relativistic quantum
field theories, which enables an analysis of the phase structure of QCD~\cite{latticeQCD}
at  finite, but imaginary, quark density on the lattice. The physics at real densities is then obtained
by analytic continuation to real values of the chemical potential. The applicability of an
analytic continuation with polynomials is restricted to small $\mu$,
more precisely $\mu/T\lesssim 1$. So far this approach has not delivered conclusive evidence for or against
the existence of a critical end-point of the line of
chiral transitions in the \mbox{$T$-$\mu$ plane}~\cite{deForcrand:2006pv}.
For non-relativistic fermions in the BEC-BCS crossover, on the other hand, the tri-critical end-point of the line of
(second-order) superfluid transitions is known to exist. Moreover, our present study suggests that this point might be
accessible in lattice calculations using complex-valued chemical potentials.
 A successful application of such an approach to ultracold Fermi gases may be
useful for future lattice QCD studies as well.
In fact, it can be very beneficial for present and future studies of the QCD phase diagram to
have an experimentally accessible system at hand which allows to test theoretical approaches and techniques in a clean and
controlled environment, especially since the experimental search for the critical point in the QCD
phase diagram has proven tremendously difficult and therefore requires reliable guidance from theory.

In this first analysis we employ a mean-field approach, as discussed elsewhere for the case
of real-valued $\mu$ (see e.g.~\cite{Holland:2001zz,revasym}), to study the phase diagram in the
complex-valued case. Although this can only be viewed as a lowest-order
approximation, it
relies only on a single input parameter {(e.g. $k_{\rm F}$)}
as is the case for the full evaluation of the associated path-integral using, e.g., MC calculations.
{Thus, our results do not suffer from a parameter ambiguity
but only from an uncertainty associated with the underlying approximation.
This can be understood on very general grounds from an analysis of the
fixed-point structure of fermionic theories~\cite{Diehl:2009ma}.
}

We begin by discussing a few general aspects of non-relativistic theories with complex-valued chemical potentials.
In general, the grand canonical partition function~$\mathcal Z$ of non-relativistic fermions reads
\be
\zgc (T,\bar{\mu},h)=\Tr\left[ \E^{-\beta(\hat{H}-\bar{\mu}(\hat{N}^{}_{\uparrow}+\hat{N}^{}_{\downarrow})
-h (\hat{N}^{}_{\uparrow}-\hat{N}^{}_{\downarrow}))}\right]\,, \label{eq:Z}
\ee
where $T$ is the temperature and $\beta=1/T$. We shall assume that the Hamiltonian~$\hat{H}$ describes
the dynamics of a theory with two fermion species, denoted by \mbox{$\uparrow$ and $\downarrow$}, interacting
via a two-body interaction. Here~$\hat{N}^{}_{\uparrow,\downarrow}$ are the particle number operators associated
with each species, and~$\mu_{\uparrow,\downarrow}$ are the corresponding chemical potentials.
For convenience we introduce the average
chemical potential~$\bar{\mu}=(\mu^{}_{\uparrow}+\mu^{}_{\downarrow})/2$
and the asymmetry parameter~$h=(\mu^{}_{\uparrow}-\mu^{}_{\downarrow})/2$.

{As is well known, MC calculations for unitary fermions can be performed}
without a sign problem for~$h\!=\!0$ {(see e.g. Refs.~\cite{Chen:2003vy,Bulgac:2005pj}).}
This is not true in general, however,
as polarization leads to a sign problem, regardless of the form of the interaction.
To proceed, we consider an imaginary-valued asymmetry parameter~$h$, corresponding to
a theory with complex-valued~$\mu^{}_{\uparrow,\downarrow}$, and
therefore define $h = \I h^{}_\text{I}$, where $h^{}_\text{I}$ is a real quantity.
It is easy to verify that MC calculations with imaginary-valued asymmetry can be studied with standard methods
without a sign problem: the fermion determinants appearing in the probability
measure are complex conjugates of one another. By analytically continuing~$\zgc (T,\bar{\mu},h^{}_\text{I})$,
one then obtains $\zgc (T,\bar{\mu},h)$, which is the central quantity in studies of imbalanced Fermi gases.

To understand whether the tri-critical end-point is
accessible with such an approach, we study the mean-field phase diagram with
complex-valued chemical potentials. We compute the mean-field potential for
the U($1$) order-parameter, starting from the path-integral representation for~$\zgc$:
\be
\zgc = \int {\mathcal D}\psi^{\dagger}{\mathcal D}\psi\, \E^{-{S} [\psi^{\dagger},\psi]
 }\,,\nn
\ee
where
\be
&& S [\psi^{\dagger},\psi]=  \int d\tau\int d^3 x\, \bigg\{ \psi^{\dagger}
\left( \partial _{\tau} - \vec{\nabla}^{\,2} -\bar{\mu} \right) \psi \nn\\
&&  \quad - h \left(  \psi^{\ast}_{\uparrow}\psi^{}_{\uparrow} -  \psi^{\ast}_{\downarrow}\psi^{}_{\downarrow}\right)
 + \, \bar{g}(\psi^{\dagger}\psi) (\psi^{\dagger}\psi)
\bigg\}\,,\label{eq:action}
\ee
and $\psi^{\rm T}=(\psi^{}_{\uparrow},\psi^{}_{\downarrow})$ and~$\bar{g}$ denotes the bare four-fermion coupling.
The dimensionless renormalized four-fermion coupling~$g\!\sim\! \bar{g}\Lambda$
is related to the scattering length~$a_{\rm s}$ by
\be
4\pi\Lambda g^{-1} =  \left( a^{-1}_{\rm s}-c_{\rm reg.}\Lambda\right).
\,
\ee
Here~$\Lambda$ denotes the ultraviolet (UV) cutoff and the constant~$c_{\rm reg.}>0$ depends on the
regularization scheme. We use units such that~$2m=1$, where~$m$ is the fermion mass.
%
\begin{figure}[t]
\includegraphics[width=\columnwidth]{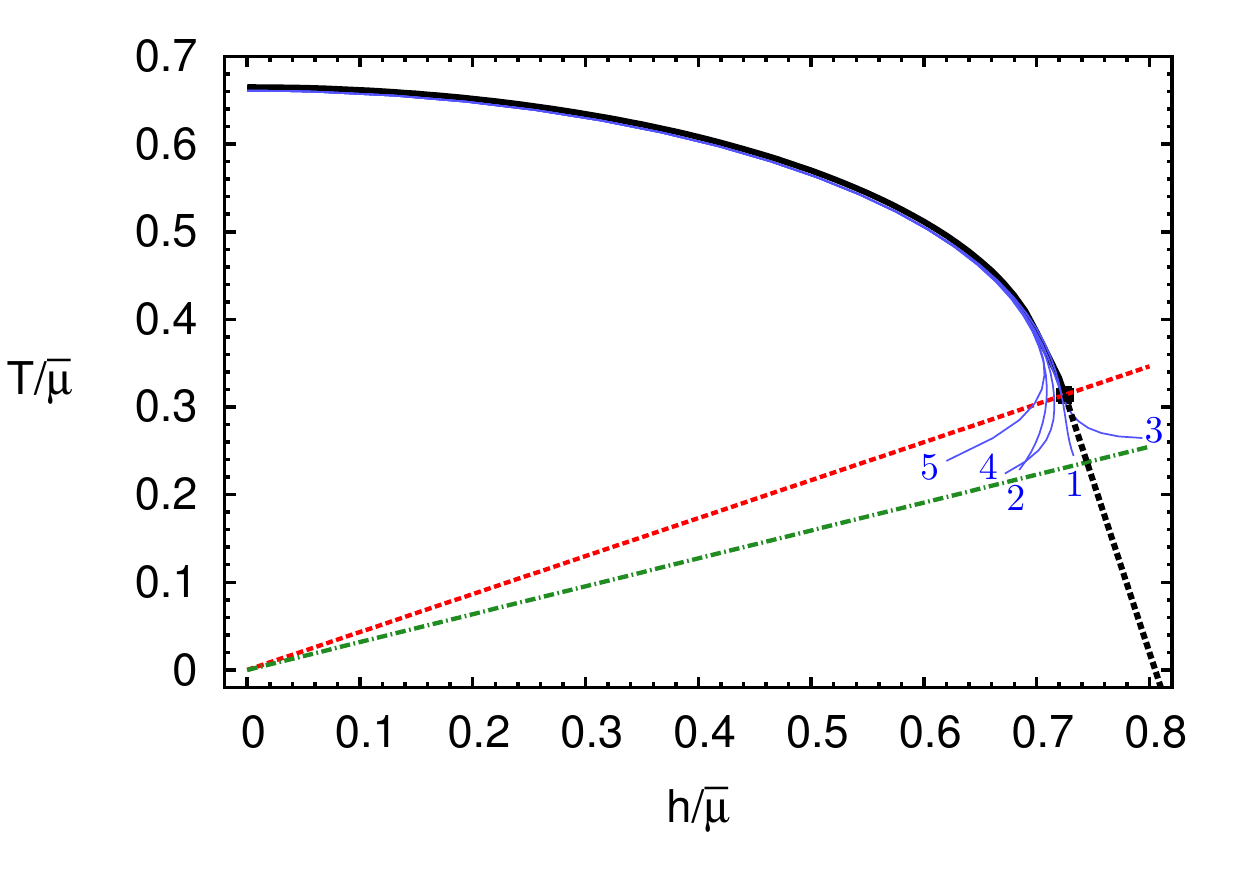}
\caption{\label{fig:1}(color online)
Phase diagram of an ultracold Fermi gas at unitarity in the ($T$,$h$) plane.
The solid (black) curve is a line of second-order phase transitions, which ends at a tri-critical point
$(h_{\rm cp}/\bar{\mu},T_{\rm cp}/\bar{\mu})$ and is followed by a line of first-order transitions
(see e.g. Refs.~\cite{revasym}). The (red) dashed line is  $T/{\bar{\mu}}=(T_{\rm cp}/h_{\rm cp})h/\bar{\mu}$ and the (green) dashed-dotted line is $\pi T/\bar{\mu}= h/\bar{\mu}$. The (light-blue) thin curves
are analytic continuations from imaginary $h$ using Pad\'{e} approximants of order $N_{\rm max}=1,2,\dots,5$ (see Eq.~(\ref{eq:pade})).}
\end{figure}
%
%
The interaction is represented by an auxiliary scalar
field~$\varphi \sim  g^{}_{\varphi}\, \psi_{\uparrow}\psi_{\downarrow}$,
where the parameter~$g^{}_{\varphi}$ is chosen to reproduce the four-fermion term in the action.
Since the resulting action is quadratic in the fermion fields, they can be integrated out. Thus, one obtains
the order-parameter potential
\be
\beta U(\varphi)&=&-2\beta\bar{\mu}|\varphi|^2
- \int \frac{d^3q}{(2\pi)^3}\ln \Bigg[ \frac{1}{2}\Bigg(
\cosh\left( \beta h\right)  \nn\\
&& \quad\; +\cosh\left(\beta {\sqrt{ (\vec{q}^{\,2} - \bar{\mu})^2 + g_{\varphi}^2 |\varphi|^2} } \right)
\Bigg)\Bigg]\,.
\label{eq:Upot}
\ee
This potential is directly related to the grand canonical potential:~$\Omega=VU(\varphi^{}_0)$, where~$V$ is the
volume of the system and~$\varphi^{}_0$ denotes the value of $\varphi$ that
minimizes the potential.
In that state, $g_{\varphi}^2 |\varphi^{}_0|^2$ can be identified with the fermion gap,~$\Delta$, the order parameter
of the spontaneously broken U($1$) symmetry, associated with a superfluid state.
From the (regularized) grand canonical potential we can obtain all thermodynamic observables.
In the unitary limit, the  dimensionless (universal) quantities, such as the critical
temperature for the superfluid transition~$T_{\rm c}/\bar{\mu}$, the corresponding gap~$\Delta/\bar{\mu}^2$ and
the ground-state energy~$E/\bar{\mu}$, are, as expected, independent of~$\bar{\mu}$ and~$g_{\varphi}$.
To compute the critical temperature~$T_{\rm c}$, it
is convenient to employ the gap equation~$(\partial U/\partial \varphi)|_{\varphi^{}_0}=0$ and
exploit the fact that the fermion gap $\Delta \sim \varphi^{2}_0$ vanishes identically at~$T=T_{\rm c}$.

\begin{figure}[t]
\includegraphics[width=\columnwidth]{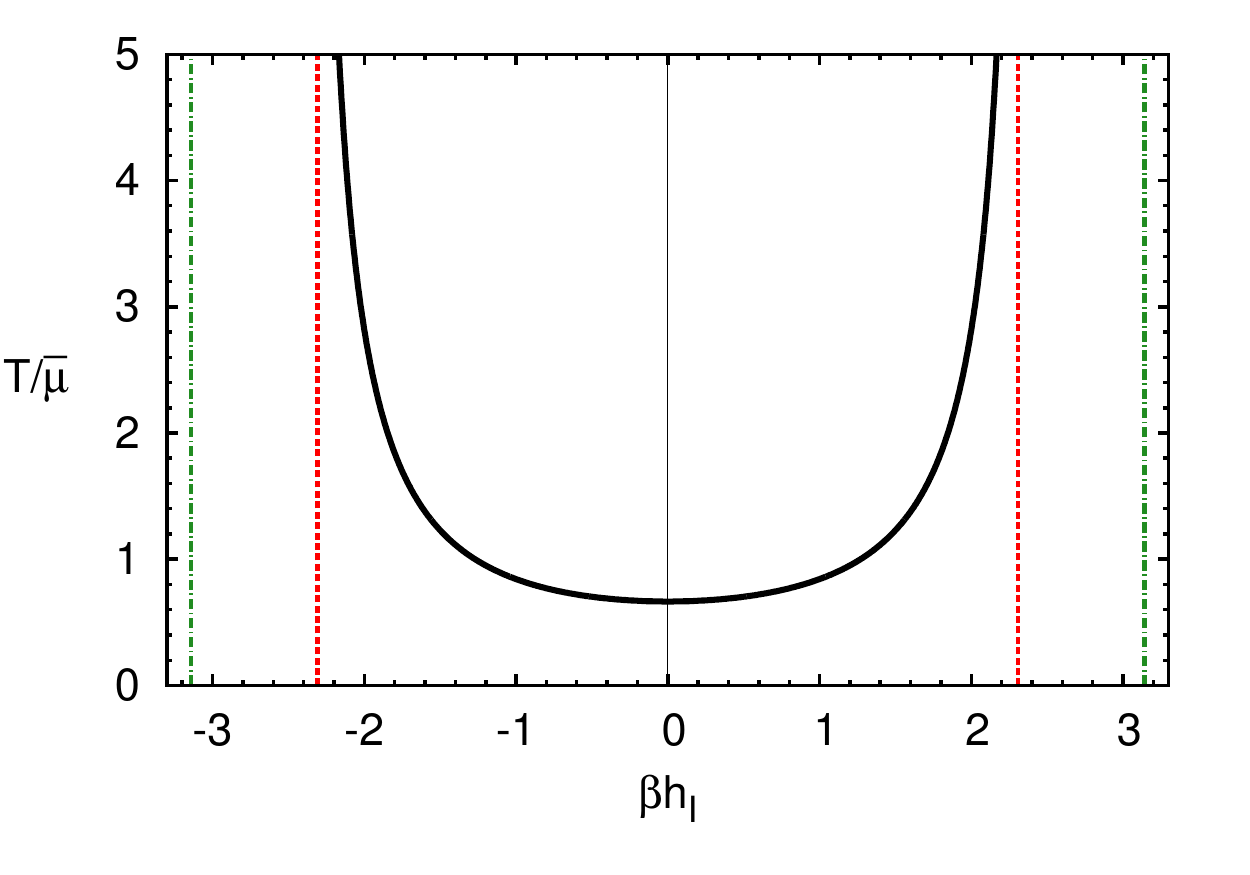}
\caption{\label{fig:2}(color online) Phase diagram in the ($T$,$h^{}_\text{I}$) plane.
The solid line is a line of second-order phase transitions below which the fermion gap is finite.
The (red) dashed and (green) dash-dotted lines are the analogs of those in Fig.~\ref{fig:1}, with $h^{}_\text{I}$ replacing $h$.
Between the (red) dashed lines and the (green) dash-dotted lines~$T_{\rm c}\to\infty$ for~$\bar{\mu}>0$.
For~$\bar{\mu}<0$, however,~$T_{\rm c}$ remains finite.}
\end{figure}

From Eq.~\eqref{eq:Upot}, it is apparent that mean-field potential $U$  is $2\pi$-periodic in
$\beta h^{}_\text{I}$.
This is a property not only of  the mean-field approximation,
but of the full theory, as can be seen by inspecting Eq.~\eqref{eq:action};
the imaginary part of~$h$ effectively shifts the Matsubara modes of the
fermions~$\nu_n=(2n+1)\pi T$ by $h_{\text I}\,T$.
[Loosely speaking,~$\partial_{\tau}$
is replaced by~$\I\nu_n$ when the action~$S$ is formulated in momentum space.]
It follows that it is not possible to study arbitrary asymmetries with this technique:
$h^{}_\text{I}$ is bounded to values~$\beta h^{}_\text{I} < \pi$.
Nevertheless, a large part of the phase diagram in the physical~$(T,h)$-plane can
be explored within this approach. As we show next, the
mean-field calculation presented here suggests that a (tri)critical point in the phase diagram may indeed be accessible
in lattice MC calculations with imaginary $h$.

In Fig.~\ref{fig:1} we show the well-known mean-field phase diagram in the $(T,h)$
plane (see e.g. Refs.~\cite{Holland:2001zz,revasym}). We refrain from discussing the appearance
of inhomogeneous phases (Sarma and/or FFLO), and focus on the phase boundaries of the
homogeneous ones. In Fig.~\ref{fig:2}, we show the corresponding phase diagram in the plane
spanned by the temperature and the imaginary-valued asymmetry parameter. As noted above, the phase
diagram is $2\pi$-periodic in~$\beta h^{}_\text{I}$. We therefore show only the domain
~$\beta h^{}_\text{I}\in [-\pi,\pi]$, bounded by the green dashed-dotted lines.

Given a phase diagram for imaginary $h$, obtained e.g. using lattice MC, one can access
only the temperature regime $T > h/\pi$ in the physical $(T,h)$ plane of Fig.~\ref{fig:1}.
Nevertheless, this represents a fairly large part of the phase diagram, which is at the heart of theoretical
and experimental studies. Most remarkably, our analysis suggests that the (tri)critical point may be located within  the
accessible part of the phase diagram, implying that this point may be within the reach of lattice MC calculations
with an imaginary asymmetry parameter. The phase transition line can then be obtained from an analytic
continuation of the results for~$T_{\rm c}(h^{}_\text{I})$, as shown below.

The phase structure of the theory in the~$(T,h^{}_\text{I})$ plane (Fig.~\ref{fig:2}) is intriguing.
{It can be shown analytically {that~$T_{\rm c}\!\to\! \infty$ for~$\beta h^{}_\text{I}=(2N+1)\pi$ with~$N \!\in\! {\mathbb Z}$
and~$\bar{\mu}>0$.}} A similar result is found in relativistic fermion models, such as
the Gross-Neveu model in~$(1+1)d$~\cite{Karbstein:2006er}.
Again, we expect this result to remain valid also beyond the mean-field approximation.
In fact, for~$\beta h^{}_\text{I} =(2N+1)\pi$ the fermionic Matsubara modes~$\nu_n=(2n+1)\pi \, T$
in~\eqref{eq:action} effectively assume the form~$\nu_n=2n\pi T$
associated with bosonic degrees of freedom.
Thus, in this case the fermions acquire a (thermal) zero mode, which tends to condense, independently of the actual
value of the temperature.
However, contrary to relativistic fermion models, we find numerically that
$T_{\rm c}\to \infty$ already for~$|\beta h^{}_\text{I}| > |(\beta h^{}_\text{I})_{\infty}| \simeq 2.397$. In other words,
 for {$|(\beta h^{}_\text{I})_{\infty}|\leq |\beta h^{}_\text{I}| <\pi$} there is
always a fermion condensate and the U($1$) symmetry is not restored by increasing the temperature.
For $|\beta h_{\rm I}| < |(\beta h^{}_\text{I})_{\infty}|$, the phase transition is
second order.
The upper  limit~$|(\beta h^{}_\text{I})_{\infty}|$ will take a different value when one goes beyond the mean-field approximation.

In our numerical studies we find that the value $|\beta_{\rm cp}h_{\rm cp}|$ associated
with the (tri)critical point
is slightly lower than $|(\beta h^{}_\text{I})_{\infty}|$. This close agreement
appears to be a mere coincidence.
In fact, in the weak-coupling limit  the difference between~$|\beta_{\rm cp}h_{\rm cp}|$
and~$|(\beta h^{}_\text{I})_{\infty}|$ is larger than in the unitary regime, at least in mean field approximation~\cite{workinprogess}.
In this context, we note that the absence of a (tri)critical point in the $(T,h_{\rm I})$ plane as well as the
absence of U($1$) restoration in the domain $|(\beta h^{}_\text{I})_{\infty}|\lesssim |\beta h^{}_\text{I}| <\pi$ does not imply their absence
for  real-valued asymmetries.

In analogy to relativistic fermion models~\cite{Karbstein:2006er},
the analytic continuation of the phase boundary reproduces the phase boundary only up to the (tri)critical point.
However, by means of an analytic continuation of the (full) order-parameter potential, the
phase diagram can be mapped out in the complete region where $\beta h < \pi$ of the physical~$(T,h)$-plane,
including the line of first-order transitions.

We stress that, using imaginary polarizations, the detection of a (tri)critical point appears to be feasible with
lattice MC calculations, even though this might require techniques for the computation the effective potential. The latter might
be borrowed from, e.~g., lattice MC studies of supersymmetric models~\cite{Wozar:2011gu}.
Also, one may employ Pad\'{e} approximants to scan the phase diagram
and detect the approximate location of the (tri)critical point
by determining the point at which the Pad\'{e} approximants do not converge
(see, e.~g., Refs.~\cite{Baker,Gupta:2011ma,workinprogess} and also our discussion below).
In any case, the analytic continuation of numerical data is difficult since one has to deal with (systematic and statistical)
uncertainties of the data from the MC calculation (see, e.~g., Ref.~\cite{Cea:2012ev}).

The grand canonical partition function~$\zgc$ is, just like the order-parameter potential~$U$, invariant under~$h \to -h$.
This allows us to expand~$\zgc$ (and other physical quantities) in powers of $(\beta h)^2$.
In the mean-field approximation, we find that the radius of convergence {for the grand canonical potential
is~$r\equiv |\beta h|_{\rm max}=\pi$ for~$\Delta\equiv 0$ and~$\bar{\mu}>0$, but~$r>\pi$ in the case of a finite
gap~$\Delta$. These observations facilitate
the analytic continuation from imaginary- to real-valued asymmetry parameter.
When performing MC calculations of ultracold Fermi gases, one now has several options for
the analytic continuation. For example, one may fit the data for an
observable~$\mathcal O$ at a given temperature~$T^{}_0 = 1/\beta_0^{}$ to the ansatz
${\mathcal O}=\sum_{n=0}^{N_{\rm max}}C^{(n)}_{\mathcal O} (\beta_0^{} h^{}_\text{I})^{2n}$.
Here $C^{(n)}_{\mathcal O}$ are constants determined by the fit to the data and $N_{\rm max}$ represents the
truncation order (whose value depends on the amount of data available). Moreover, it is assumed that~$\mathcal O$
has been made dimensionless with, e.g., a suitably chosen power of~$\bar{\mu}$. By means of a simple analytic continuation
of the polynomial, one then obtains the dependence of~$\mathcal O$ on~$h$.

Within the mean-field approximation, we find that the pressure for $0 \leq \beta_0^{}h \lesssim 1$
(at a temperature $T_0 = 1/\beta_0^{}\approx \bar{\mu}/2$) can, to a good approximation, be recovered
from a fit to the imaginary-$h$ data with~$N_{\max}=2$. Given an approximation for the pressure, one can in principle compute
the energy as a function of~$h$. At zero temperature, this would be equivalent to knowing the~$h$-dependence of
the so-called {\it Bertsch} parameter. However,
zero-temperature values of physical observables for finite~$h/{\bar\mu}$ are obviously
not directly accessible within such an approach. Nonetheless, it is known from lattice MC calculations that below the superfluid
transition the {\it Bertsch} parameter at~$h=0$ rapidly approaches its zero-temperature value~\cite{Bulgac:2005pj}.
In mean-field approximation we find a similar behavior, also at finite~$h/\bar{\mu}$. It is therefore
conceivable that a reliable estimate of the {\it Bertsch} parameter at~$T=0$ and finite polarization can be extracted from
lattice calculations at finite temperatures and imaginary~$h$ by means of an analytic continuation.

One may perform the analytic continuation using more elaborate fit functions such as
Pad\'{e} approximants, also used in lattice QCD studies~\cite{latticeQCD}. In Fig.~\ref{fig:1}, for example, we
have reconstructed the phase boundary at real-valued asymmetry by  fitting the phase transition line
in the $(\beta h_{\rm I},\beta\bar{\mu})$-plane with the function
\be
\label{eq:pade}
C\frac{1+\sum_{i=1}^{N_{\rm max}}a_i [1-\cos(\beta h_{\rm I})]^i}{1+\sum_{j=1}^{N_{\rm max}}b_j [1-\cos(\beta h_{\rm I})]^j}\,,
\ee
where~$N_{\rm max}$ again defines the truncation order. The coefficients $a_i$'s, $b_i$'s and the constant~$C$ are determined by the
fit. This ansatz respects the $2\pi$-periodicity in $\beta h_{\rm I}$ and
can be generalized to observables other than~$T_{\rm c}$.
In Fig.~\ref{fig:1} we show the results for the critical temperature~$T_{\rm c}$ obtained using such a fit for~$N_{\rm max}=1,2,\dots,5$, see Ref.~\cite{workinprogess}
for details. Finally, we note that the fits may be even further optimized by choosing even more elaborate
sets of basis functions~\cite{Skokov:2010uc}.

We have completely disregarded any discussion of inhomogeneous phases and, in particular, the possible existence of such phases in
the $(T,h^{}_\text{I})$ plane. While such a discussion is left to future work, we do not expect
an inhomogeneous condensate~$\varphi^{}_0 (\vec{x})\sim {\rm e}^{{\rm i}\vec{q}_0\cdot \vec{x}}$ to show up for imaginary~$h$. In its
well-known form, the (center-of-mass) momentum $\vec{q}_0$ is determined by the difference in the chemical
potentials of the spin-up and spin-down fermions. From a naive point of view,
one expects that the solutions $\varphi^{}_0(\vec{x})$ of the quantum equation of motion turns
into~$\varphi^{}_0 (\vec{x})\sim {\rm e}^{-\vec{q}_0\cdot \vec{x}}$ for complex-valued chemical potentials
and hence no longer define the ground state.

We have discussed the possibility of studying polarized Fermi gases
with the aid of complex-valued chemical potentials. While the latter are
not required in analytic studies, they are in MC calculations which otherwise would suffer from the sign problem.
We have argued that the (tri)critical point is in principle within reach in this framework
and that the zero-temperature limit of observables might be indirectly accessible as well.
This work therefore
suggests that, together with the experimental data at hand, future {\it ab initio}
MC calculations with complex-valued chemical potentials have the capacity to
push our understanding of collective many-body phenomena to a new level.
Our present study marks the starting point and can already be used to guide
these calculations.

JB would like to thank M. Buballa, F. Karbstein, O. Philipsen, and D. Rischke for useful discussions.
JB acknowledges support by the DFG under Grant BR 4005/2-1
and by the Helmholtz International Center for FAIR within the LOEWE program of the State of Hesse.
JWC is supported by the NSC(99-2112-M-002-010-MY3) of ROC and CASTS \& CTS of NTU.
BF is supported in part by the ExtreMe Matter Institute EMMI.

\bibliographystyle{h-physrev3}

\end{document}